\documentstyle[epsfig]{mn2e}
\input{psfig.sty}
\newif\ifAMStwofonts
\def\Mesz{M\'esz\'aros~}
\def\Pacz{Paczy\'nski~}

\title[Gamma-ray bursts and star-forming galaxies]{Gamma-ray
bursts in normal and extreme star-forming galaxies}

\author[N. Trentham, E. Ramirez-Ruiz and A. W. Blain]
{Neil Trentham,$^1$ Enrico Ramirez-Ruiz$^1$ and A.~W.~Blain$^{1,2}$\\
$^1$ Institute of Astronomy, University of Cambridge, Madingley Road,
CB3 0HA Cambridge, UK \\
$^2$ Astronomy Department, Caltech 105-24, Pasadena, CA 91125, USA}

\begin{document}
\maketitle

\begin{abstract} 
{ 
We discuss how gamma-ray burst
(GRB) optical afterglows and multiwavelength  
observations of their host
galaxies can be used to obtain information about the relative
amounts of star formation happening in optical and submillimetre 
galaxies.  
That such an analysis will be possible
follows from the currently-favoured
idea that GRBs are closely linked with high-mass star formation. 
Studying GRB host galaxies offers a method of finding low-luminosity 
submillimetre galaxies, which  
cannot be identified either in optical Lyman break surveys, because so 
much of their star formation is hidden by 
dust, or in 
submillimetre surveys, because their submillimetre
fluxes are close to or below the confusion limit. 
Much of the star formation in the Universe could have
occurred in such objects, so searching for them is
an important exercise. 
From current observations, GRB host
galaxies appear to be neither 
optically-luminous Class-2 SCUBA galaxies like SMM\,J02399$-$0136 or 
SMM\,J14011+0252, 
nor galaxies containing dense
molecular cores like local ultraluminous infrared galaxies (ULIGs), but 
rather some
intermediate kind of galaxy.    
The host galaxy of GRB 980703 is a prototype of this kind of galaxy.
}
\end{abstract} 


\section{Introduction}
\label{sec:int}

Much of the star-formation activity
happening in the Universe at redshifts $z>1$ is
hidden by dust (e.g.~Blain et al.\ 1999a,b, 2002; Chary \& Elbaz 2001).
At low redshift ($z \ll 1$), the amount of star formation in
optical/ultraviolet (UV)
galaxies accounts for almost all of the star
formation happening in the Universe, 
but at higher redshift it could well be much lower,
due to the existence of many
heavily dust-enshrouded submillimetre galaxies.

The populations of galaxies that are bright at optical/UV and
submillimetre wavelengths 
are largely disjoint.  The optical/UV galaxies
which dominate at high-$z$ are Lyman-break galaxies (Steidel et al.~1999)
with typical 850-$\mu$m submillimetre 
fluxes of about 0.1\,mJy (Peacock et al.~2000).  
The submillimetre 
galaxies which are most common at high-$z$ 
tend to have optical
magnitudes $\, I>26 \,$ (Smail et al.~2002).  
The reason that these galaxies are so faint is presumably due to internal
dust extinction removing most of the emitted rest-frame ultraviolet light, 
which corresponds to observed-frame optical light.  This internal
extinction is similarly important in the most luminous dust-enshrouded
galaxies locally, the ULIGs (Sanders \&
Mirabel 1996), which are perhaps
the most plausible local counterparts to the
high-$z$ submillimetre galaxies, but tend to have
very low ultraviolet fluxes (Goldader et al.~2002).  

Despite what is written in the previous paragraph, a few of the
best-studied high-$z$
submillimetre galaxies have properties quite unilke 
faint field galaxies with $I>26$.
These include 
\begin{enumerate} 
\item Class-2 SCUBA sources (Ivison et al.\ 2000; Smail et al.~2002) like
SMM\,J02399$-$0136 (Ivison et al.~1998) and SMM\,J14011+0252 (Ivison et al.\
2001), which are
extremely luminous at optical wavelengths and have rest-frame
intrinsic $B$ magnitudes more than one magnitude
brighter than any
of the luminous Lyman-break galaxies studied 
by Pettini et al.~(2001). 
The star-formation rates of Class 2 SCUBA
sources implied by both rest-frame $B$-band
luminosities and H$\alpha$ luminosities are high, 
comparable to those derived from
the submillimetre luminosities.  
The star formation in these sources could be only lightly obscured, 
leading to 
the bright optical fluxes; 
\item Extremely red objects (EROs), or Class-1 SCUBA sources 
(Smail et al.\ 2002). These 
could be more  
obscured versions of the
Class 2s, in which most rest-frame optical light escapes but not most
rest-frame ultraviolet light.  
\end{enumerate} 
These two types of object are easy to study in detail because optical or
near-infrared identifications are readily available, but most
evidence (Smail et al.~2002) suggests that they do not represent
the bulk of the submillimetre galaxy population.
Indeed, the difficulty in
obtaining unambiguous optical identifications is one of the most
important hindrances to our understanding the submillimetre
population.

A current observational problem is that only the brightest submillimetre
galaxies can be found in direct surveys (using the SCUBA camera 
on the James Clerk Maxwell Telescope; Holland et al.~1999)
due to confusion at 850-$\mu$m fluxes below 
$\approx$ 2\,mJy (Blain, Ivison \& Smail 1998).  Yet much of the star formation
in the Universe could be happening in submillimetre galaxies  
with fainter fluxes, given the constraints
jointly imposed by the submillimetre and far-infrared backgrounds
and source counts (Blain et al.~1999a,b).     
These objects cannot be found easily in 
optical field surveys (e.g.~Steidel et al.~1999)  
either,
since their optical fluxes are low (Ivison et al 2000; Smail et al.~2002). 
This situation may change at low redshifts ($z \leq 2$) 
in the near future however, with the advent of
deep mid- and
far-infrared field surveys from the {\it SIRTF} 
satellite 
(e.g.~Lonsdale 2001).

In a separate development, evidence is now accumulating
that gamma-ray bursts (GRBs) are intimately linked with the deaths
of young massive stars.
This is because (i) a natural 
mechanism for producing a long-duration ($>$ 2 s)
burst is the cataclysmic collapse
of a massive star leading to a hypernova-type explosion
(Woosley 1993; Paczy{\'{n}}ski 1998;  MacFadyen \& Woosley 1999; MacFadyen,
Woosley \& Heger 2001);
(ii) the GRB host galaxies which have been
identified often show signs of ongoing star formation, such as
very blue colours and prominent emission lines (Bloom et al.~1998a;
Kulkarni et al.~1998), high radio continuum fluxes (Berger, Kulkarni \& Frail
2001), and high submillimetre fluxes (Frail et al.~2002); 
(iii) iron line and edge features  
have been identified in two bursts (Piro et al.~2000; Amati et al.~2000),
suggesting a link with supernovae.  The mass of iron implied by
these measurements is uncertain but consistent with the masses of iron 
produced in core-collapse supernovae;  
(iv) in at least four bursts a red component
with a flux consistent with that of a supernova was discovered several
weeks after the burst (Bloom et al.~1999a; Reichart 1999; Bj\"ornsson
et al. 2001;  Lazzati, Covino \& Ghisellini 2002).   
GRBs are therefore thought to be found in regions of ongoing
star formation, since young massive stars do not live long enough to move
significantly from the sites of their birth.  

An afterglow at rest-frame optical wavelengths has been observed following
many GRBs,  
thought to be due to the interaction between the relativistic blast wave 
originating from the central engine and the local interstellar medium
(Rees \&  \Mesz 1992, 1994; \Pacz \& Xu 1994; Katz 1994; Sari \& Piran 
1995; \Mesz \& Rees 1997).  Whether or not this
optical afterglow  
reaches the observer is a function of the line-of-sight
extinction towards the GRB (Waxman
\& Draine 2000; Venemans \& Blain 2001; Draine \& Hao 2001;
Fruchter, Krolik \& Rhoads 2001b, Reichart \& Price 2002), 
which is likely to be very different for bursts 
happening within submillimetre and optical/UV galaxies.

The line-of-sight extinction to a GRB is, however, complicated by local
physical processes.  For example, optical/UV/X-ray radiation from the GRB
may heat and sublime nearby dust (Waxman \& Draine 2000,
Galama \& Wijers 2001, Venemans \& Blain
2001, Fruchter et al.~2001b).  But radiation from the GRB
is unlikely to destroy the dust responsible for the bulk of the
extinction in submillimetre galaxies
since the X-ray/UV flux scales with distance $d$ as $\sim d^{-2}$ so that dust
grains at $\gg 10$\,pc from the GRB are unaffected, and
it is dust on these large scales that is 
responsible for most of the internal extinction.
On the other hand, Fruchter et
al.~(2001b) recently suggested that the dust grains 
most responsible for internal extinction 
may indeed be destroyed 
by strongly beamed X-ray radiation if  
the electrostatic stresses generated by
the X-rays exceed the grain tensile
strengths.  

If most submillimetre  
galaxies are optically faint due to huge amounts of internal extinction, 
then we expect the optical
afterglow of most GRBs occurring within them to be completely
extinguished.  This would 
certainly happen were most submillimetre galaxies to be
similar to local ULIGs like Arp 220 
since there are typically tens of magnitudes of 
visual extinction to their 
cores (e.g.~see Scoville et al.~1998). 
Indeed, Ramirez-Ruiz, Trentham \& Blain
(2001) suggest this as the reason for undetected optical afterglows
in about half the GRBs studied by Lazzati et al.~(2002).   

Here we assess observations of optical afterglows 
in conjunction with submillimetre
limits to the flux
densities of GRB host galaxies.
Of particular interest is the possibility of using GRB host galaxies
to find low-luminosity submillimetre galaxies which cannot be found in
either SCUBA or optical surveys.
Interpretations based on current data and prospects for the future
are presented.  
Throughout the paper we assume 
$H_0 = 65\,\, {\rm km} \, {\rm s}^{-1}
\, {\rm Mpc}^{-1}$,
$\Omega_{\rm m}=0.3$,
and $\Omega_{\Lambda}=0.7$.  

\section{Observations of GRBs in submillimetre galaxies}

\subsection{Submillimetre observations and the host of GRB\,010222}

Most current evidence suggests that
the fraction of cosmic star formation happening in submillimetre galaxies 
$f_{\rm submm}$ is
high, with estimates ranging up to about 0.9. 
The value of $f_{\rm submm}$ might
however be poorly constrained by 
current observations due to uncertainty
in the form of both evolution of the luminosity function of
infrared galaxies and their spectral energy 
distributions (SEDs) -- compare the results of 
Blain et al. (1999a,b);
Trentham, Blain \& Goldader 1999, Eales et al.~(1999, 2000) 
and Gispert, Lagache \& Puget (2000). 
One reason that these functions are relatively poorly
constrained is that no redshifts are currently known for 
high-redshift submillimetre galaxies, except for a small number
of Class-2 SCUBA galaxies (Smail et al.~2002), which do not comprise 
the bulk of the submillimetre galaxy population, and for 
the host galaxies of GRB\,010222 and 
GRB\,000418 (see below).  Another reason is that
the dust temperatures of the submillimetre galaxies
is unknown (it has only been measured for a small number
of Class-2 SCUBA galaxies), and the bolometric
luminosity of a submillimetre galaxy is a strong
function of the dust temperature.
Other uncertainties,
like the conversion from infrared luminosities to
star-formation rates, reflecting uncertainties in the 
initial stellar
mass function are potentially important too. 

Noting the above, in this work we
follow Ramirez-Ruiz al.~(2001) and estimate 
$f_{\rm submm} = 0.82 \pm 0.10$ (based on the models of
Blain et al.\ 1999a,b). 
Within the submillimetre galaxy population,
a fraction 0.10 -- 0.20 of the total
star formation happens in sources with observed 850-$\mu$m fluxes 
$S_{850} > 4$\,mJy
(see Fig.~7 of Ramirez-Ruiz et al.~2001).
This transition value
of 4 mJy is chosen because
it represents an attainable flux threshold for detecting GRB host
galaxies, given several hours of observing with SCUBA.
Star-formation integrated over cosmic time is 
then partitioned between the various type of galaxies as follows: 
SCUBA galaxies with $S_{850}$ $>$ 4 mJy: 0.07 -- 0.18 ; 
SCUBA galaxies with $S_{850}$ $<$ 4 mJy: 0.58 -- 0.83;
UV galaxies: 0.08 -- 0.28.
If GRBs are associated with recent star formation, as outlined in Section 1,
then they should happen in the three different types of galaxies in these 
proportions since they can be seen at all the redshifts.

The most complete survey for submillimetre host galaxies
of GRBs with afterglows using SCUBA is that 
of Smith et al.~(1999, 2001).  Their results are presented in Table 1. 
In a separate study,  
GRB\,010222 ($z=1.476$) was found to have a time-independent
850-$\mu$m SCUBA detection of
$\sim 4$\,mJy, which is likely to
originate from the host galaxy (Frail et al.~2002).
Interestingly, this burst had a bright optical afterglow, meaning
that it did not happen in a region of very heavy obscuration.
Yet this is not a Class-2 SCUBA galaxy since its 
optical afterglow light curve (Holland et al.~2001a) does not flatten at
$R \sim 22$. Recently, Berger et al.\ (2002) detected submillimetre 
and radio emission consistent with the host galaxy of GRB\,000418. 

\begin{table}
\caption{SCUBA measurements for the hosts of GRBs with optical afterglows. 
The first set of entries are the 3$\sigma$ limits from Smith et al.~(1999, 2001).
The second entries are the measured 850-$\mu$m flux of the hosts
of GRB\,010222 and GRB\,000418 from
Frail et al.~(2002) and Berger et al.~(2002).  
The third entry is the 850-$\mu$m flux
inferred for
the host of
GRB 980703 from the radio measurements of Berger et al.~(2001).}
{\vskip 0.75mm}
{$$\vbox{
\halign {\hfil #\hfil && \quad \hfil #\hfil \cr
\noalign{\hrule \medskip}
GRB & $z$ & 850-$\mu$m flux &\cr
\noalign{\smallskip \hrule \smallskip}
970508 & 0.84 & $<$ 30 mJy &\cr
971214 & 3.42 & $<$ 3.04 mJy &\cr
980326 & unknown & $<$ 8.1 mJy  &\cr
980329 &$<$3.9& $<$ 3.5 mJy  &\cr
980519 & unknown & $<$ 7.3 mJy  &\cr
980703 & 0.97 & $<$ 5.4 mJy  &\cr
981220 & unknown & $<$ 4.9 mJy  &\cr
981226 & unknown & $<$ 12 mJy   &\cr
991208  & 0.71 & $<$ 4.9 mJy  &\cr
991216  & 1.0  & $<$ 3.5 mJy  &\cr
000301C & 2.03 & $<$ 3.6 mJy  &\cr
000630  & unknown & $<$ 6 mJy  &\cr
000911  & 1.06 & $<$ 2.6 mJy  &\cr
000926  & 2.04 & $<$ 20 mJy  &\cr
\noalign{\smallskip \hrule \smallskip}
010222 & 1.48 & 3.7 $\pm$ 0.53 mJy  &\cr
000418 & 1.18 & 3.6 $\pm$ 0.8 mJy &\cr 
\noalign{\smallskip \hrule \smallskip}
980703 & 0.97 & 1.9 mJy  &\cr
\noalign{\smallskip \hrule \smallskip}
\noalign{\smallskip}\cr}}$$}
\end{table}

Only five bursts in the Smith et al.~(1999)
sample have 3$\sigma$ flux limits below 4 mJy.
This suggests that the fraction of cosmic star
formation happening in submillimetre galaxies
with $S_{850} > 4$ m Jy is less than 0.2.  
Even if we arbitrarily include GRB\,010222 and GRB\,000418 (where
we have {\it a priori} knowledge of a submillimetre
detection; see 
Table 1),
this fraction is only about 0.28. 
From the numbers above, 
we expect 0.07 -- 0.18 of
the total cosmic star formation to have happened in
submillimetre galaxies
with $S_{850} > 4$ mJy, so this is consistent
with the SCUBA results.  At present the statistics
from SCUBA are poor, but a moderate increase (a
factor of a few) in the number of bursts observed
could allow strong constraints to be placed (see Barnard et al.\ 2002).

There is however one 
potentially serious complication. 
All the bursts in Table 1 have detected optical afterglows.
Indeed, this is perhaps the best way of determining
an accurate position to allow
submillimetre measurements
to be made.
Whether or not the optical afterglows
from GRBs emerge from the host galaxies and
are visible to a distant observer depends
on the line-of-sight internal extinction.  For UV and Class-2 SCUBA
galaxies, most optical afterglows will emerge.  For most
submillimetre galaxies (including EROs),
optical afterglows
are expected to emerge only if they happen in a region of moderate
obscuration.  They are unlikely to emerge from well within compact
molecular cores (Ramirez-Ruiz et al.~2001; Venemans
\& Blain 2001), which is the type of environment that most
star formation is happening in local ULIGs
(Condon et al.~1991;
Surace et al.~1998; Surace \& Sanders 1999; Soifer et al.~1999).

What this means is that bursts in submillimetre galaxies may
be selected against in SCUBA-observed samples since these could  
be the ones without optical afterglows.  On the other hand, two
lines of evidence suggest that this might not be a serious
concern after all.  First, the three bursts known to have occurred in
submillimetre galaxies (GRB\,010222, GRB\,000418 and GRB\,980703) all {\it did}
have bright optical afterglows.  Secondly, bursts where the optical
transient has been extinguished due to internal extinction will
still have radio afterglows, and bursts with radio but
no optical afterglows seem to be extremely rare: only two are known 
(GRB\,990506 and GRB\,970828). 

\subsection{Radio observations of the  
GRB\,980703 host galaxy}

The host galaxy of GRB 980703 ($z=0.97$) has
a high star-formation rate of about $500$\,M$_{\odot}$\,yr$^{-1}$,
inferred from its radio continuum luminosity (Berger et al.~2001),
but only 10--30\,M$_{\odot}$\,yr$^{-1}$
inferred from optical measurements (Djorgovski
et al.~1998).
The reason for this disconcordance is likely to be that
most of the star formation responsible for the high
radio flux is happening in optically thick regions.
Yet this burst had an optical afterglow, perhaps suggesting, like
GRB\,010222 that while most of the star formation
is hidden by dust, the obscuration is not {\it so} severe that
high-intensity GRB afterglows are wiped
out.  Again, this is not a Class-2 SCUBA galaxy, since the very bright
optical afterglow (Holland
et al.~2001b) did not plateau at $R \simeq 22$ at late times.

The luminosity of this host is not high enough 
for the 850-$\mu$m flux to exceed the limit 
measured by 
Smith et al.~(1999).  Berger et al.~(2001)
estimate $S_{850}=1.9$\,mJy (from the
radio--far-infrared 
correlation.
This is the only known example of a high-redshift galaxy
with a submillimetre flux $S_{850}$ $<$ 4 mJy
with an optical counterpart and a redshift measurement. 
 
\section{GRB hosts as low-luminosity submillimetre galaxies}

In the previous sections we have hypothesized that much of  
the star formation in the Universe happened in submillimetre
galaxies with $S_{850} < 4$ mJy, that such galaxies will be
missing from both optical and submillimetre surveys, and that they
may be found in samples of GRB host galaxies. They may indeed be very
common in GRB host galaxy samples. If more than half the
star formation in the Universe happened in such objects, then
the majority of GRB host galaxies may be of this type.  The fact
that the host galaxy of GRB\,980703, one of the first to be
studied in detail at non-optical wavelengths is 
such an object gives some credibility to this speculation.  

We now investigate this possibility
and discuss how to compile such a sample.
Deep optical observations of about thirty GRB host
galaxies exist, many of which would be low-luminosity submillimetre
galaxies in the context of this discussion. 
We suggest some possible
candidates.  We then outline how both
submillimetre and deep radio measurements of
GRB hosts, like those of the GRB\,980703 host
(Berger et al.~2001), could be used to verify if any  
such candidates really are low-luminosity submillimetre
galaxies.  We also discuss what it could mean if
submillimetre
galaxies are {\it not} represented in such samples.
Finally we discuss the extent of the dust obscuration
in low-luminosity submillimetre
galaxies and demonstrate that in the current scenario
they cannot be either Class-2 SCUBA
galaxies or high-redshift analogues of local ULIGs. 
 
\subsection {Optical measurements of host galaxies} 

Optical observations are inefficient at revealing submillimetre
galaxies.  The reason for this is that internal extinction
in submillimetre galaxies is large so that it is possible for
them to have very high infrared and bolometric luminosities but low
optical flux densities (Rieke \& Lebofsky 1986; Soifer et al.~1987).

Optical observations {\it can} however be used to
reveal dust-enshrouded star formation if the extinction 
is only weakly optically thick, and so the redness of a galaxy
at rest-frame ultraviolet wavelengths is a measure of
the amount of extinguished star formation
(Meurer 
et al.~1995; Meurer, Heckman \& Calzetti
1999).  However
this is not the kind of extinction found in the more
luminous submillimetre galaxies (Goldader et al.~2002).  For
example, the huge amount of hidden star formation in the
host of GRB 980703 was not revealed by optical (rest-frame
ultraviolet) measurements (Chary, Becklin \& Armus 2002). 
Presumably this is because this galaxy is reasonably 
thick, so the Meurer relation will not work. 
This host could well represent a 
typical case, meaning that optical observations 
of host galaxies will be of limited use in finding
submillimetre galaxies. 
Nevertheless, a great deal of optical information about GRB
host galaxies exists, and it is worth examining this data to  
see if we can find any clues about the nature of the galaxies.  
In Fig.\,1 we present $R$ magnitudes and redshifts for all bursts where
measurements have been made. 
For the 15 bursts (not including GRB\,010222) with both known
redshifts in the range $0.5 < z < 2.5$, which 
are likely to contribute most to 
total cosmic star formation and with host galaxy detections, 
2 have host galaxies with
$R<23$, 10 have $23.5<R<27$, and 
3 have $R>27.5$. 
Immediately,
this  tells us that most of these bursts are not happening in
Class-2 SCUBA galaxies, since their host galaxy $R$ magnitudes are too faint.  

\begin{figure*}
\psfig{file=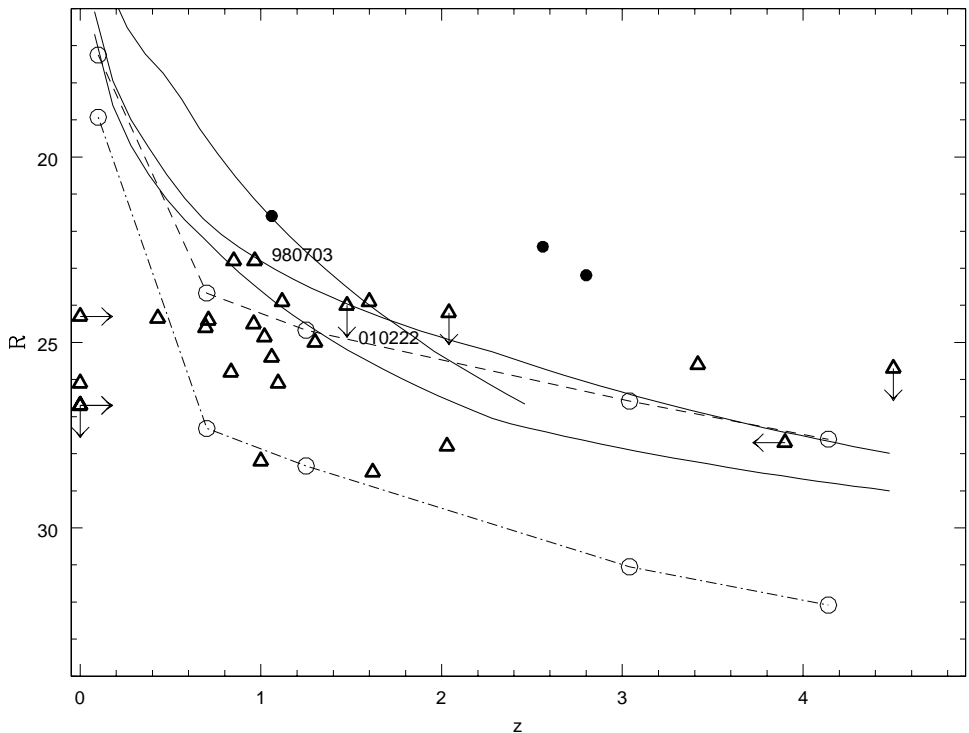,angle=-90,width=0.80\textwidth}
\caption{{Optical $R$-band properties of GRB host galaxies, submillimetre
galaxies and UV-bright galaxies.
The triangles represent the host galaxy redshifts and magnitudes
for the sample of bursts compiled in Table 2. 
The solid lines represent the three local ULIGs studied at ultraviolet
wavelengths by Trentham et al.~(1999) as they would be seen 
at different redshifts.  The three filled circles represent the 
magnitudes of Class-2
SCUBA galaxies from Smail et al.~(2002) with 
spectroscopic redshifts,
corrected for the effects of lensing.  
The open circles represent UV-bright
galaxies of luminosity $L'$, where
$\int_{L'}^{\infty} L \phi(L) \, {\rm d}L / 
\int_0^{\infty} L \phi(L) \, {\rm d}L $
= 0.5 (upper) or 0.9 (lower), at different redshifts.
The luminosity function $\phi(L)$ was
computed from the Las Campanas Redshift Survey (Lin et al 1996) at
$z=0$, from 
the survey of Cowie et al. (1999) at $z=0.7$ and $z=1.25$,
and from the 
Lyman-break sample of Steidel et al.~(1999)
at $z=3$ and $z=4$.  For all samples, the template Scd SED
of Coleman et al.~(1980) was used to convert
optical luminosities to observer-frame Kron-Cousins $R$-band magnitudes.
}
\label{fig:rfig}}
\end{figure*}

\begin{table*}
\caption{Host galaxy magnitudes for GRBs, compiled and updated from
Holland et al.~(2000a) and Djorgovski (2001a,c). 
} {\vskip 0.75mm} {$$\vbox{
\halign {\hfil #\hfil && \quad \hfil #\hfil \cr
\noalign{\hrule \medskip}
Burst & Redshift (how measured) & Host $R$ mag. & Reference  &\cr
\noalign{\smallskip \hrule \smallskip}
\cr
GRB 010222 & 1.48 (Mg II + Fe II abs.~) & $>24$   &
Djorgovski et al.~(2001a) and references therein &\cr
GRB 000926$^{*}$ & 2.04 (6 strong abs.~lines including Ly$\alpha$) & $>24.2$   &
Rol et al.~(2000) &\cr
GRB 000911$^{*}$ & 1.06 (host emission) & $ 25.4$   &
Lazzati et al.~(2001) &\cr
GRB 000630$^{*}$ &      & $ 26.7$   &
Djorgovski et al.~(2001a) &\cr
GRB 000418 & 1.12 (O[II] host emission) & $ 23.9$   &
Fruchter \& Metzger (2001) &\cr
GRB 000301C$^{*}$ & 2.03 (varous weak abs.~lines) & $ 27.8$   &
Fruchter \& Vreeswijk (2001) &\cr
GRB 000131 & 4.50 (Lyman break) & $>25.7$   &
Andersen et al.~(2000) &\cr
GRB 991216$^{*}$ & 1.02 (Fe II + Mg I + Mg II abs.~) & $24.85$   &
Djorgovski et al.~(2001a) and references therein &\cr
GRB 991208$^{*}$ & 0.71 (O[II] + O[III] host emission) & $24.4$   &
Djorgovski et al.~(2001a) and references therein &\cr
GRB 990712 & 0.43 (Mg II/I abs., O[II]/[III]+H$\gamma$/$\beta$
emiss.~) & $24.35$   &
Fruchter et al.~(2000b) &\cr
GRB 990705 & 0.85 (transient X-ray abs.~edge) & $22.8$   &
Djorgovski et al.~(2001a) and references therein &\cr
GRB 990510 & 1.62 (Mg II + Fe II abs.~) & $28.5$   &
Djorgovski et al.~(2001a) and references therein &\cr
GRB 990506 & 1.30 (see Bloom, Frail \& Sari 2001b) & $25.0$   &
Holland et al.~(2000b) &\cr
GRB 990308 &      & $>26.7$   &
Holland et al.~(2000c) &\cr
GRB 990123 & 1.60 (Mg II + Fe II abs.~) & $23.9$   &
Djorgovski et al.~(2001a) and references therein &\cr
GRB 981226$^{*}$ &      & $24.30$   &
Holland et al.~(2000d) &\cr
GRB 980703$^{*}$ & 0.97 (O[II] host emission) & $22.8$   &
Bloom \& Kulkarni (2000) &\cr
GRB 980613 & 1.10 (O[II] host emission) & $26.1$   &
Holland et al.~(2000e) &\cr
GRB 980519$^{*}$ &      & $26.1$   &
Bloom et al.~(1998b) &\cr
GRB 980329$^{*}$ & $<3.9$ & $27.7$   &
Holland et al.~(2000f) &\cr
GRB 980326$^{*}$ & $\sim 1$ (putative supernova)
& $28.2$   &
Fruchter et al.~(2001a) assuming $V-R=1$ &\cr
GRB 9712l4$^{*}$ & 3.42 (Ly$\alpha$ host emiss.~+ OI abs.~) & $25.6$   &
Odewahn et al.~(1998), Djorgovski et al.~(2001a)&\cr
GRB 970828 & 0.96 (host emission lines) & $24.5$   &
Djorgovski et al.~(2001a,b) &\cr
GRB 970508$^{*}$ & 0.835 (abs.~+ emiss.~lines) & $25.8$   &
Castro-Tirado \& Goroabel (1998) &\cr
GRB 970228 & 0.70 (host emission lines) & $24.6$   &
Bloom et al.~(2001a) &\cr
\noalign{\smallskip \hrule}
\noalign{\smallskip}\cr}}$$}
\begin{list}{}{}
\item[$^{\mathrm{*}}$]Observed with SCUBA by Smith et al.~(1999, 2001);
see Table 1
\end{list}
\end{table*}

We studied optical images of
each of the host galaxies in Table 2 and suggest that the
following are the best few candidates for being 
submillimetre galaxies with $S_{850}$ $<$ 4 mJy. 
Our assessments
are based on a number of indirect
considerations e.g.~similarity to the  
host galaxy of GRB 980703 
and morphology (spiral structure, as in the case of the host
of GRB 990705, suggests
a host galaxy is a UV-bright galaxy as opposed to a
submillimetre galaxy).

\subsubsection{GRB\,000301C}

{\it HST}-STIS imaging shows this host galaxy (Fruchter \& Vreeswijk 2001)
to be compact and only
marginally resolved. The 
FWHM resolution of the {\it HST} image 
corresponds to about
1\,kpc at the redshift of this galaxy ($z=2.03$). 

\subsubsection{GRB\,991208}

In {\it HST}-STIS images
this host galaxy appears to be compact
and only
marginally resolved (Fruchter et al.~2000a), with a FWHM of only
730\,pc.  
Furthermore, a $K$-band source (Bloom et al.~1999b) was
observed coincident
with this burst one week later.  This could be a near-infrared afterglow,
although Bloom et al. note that this would then imply an
optical--near-infrared spectral index inconsistent with the spectral
index derived from optical spectroscopy.   An alternative explanation
is that the $K$-band light comes from an ERO component. 

\subsubsection{GRB\,990506}

This burst came from a compact galaxy with $r_h=0.94$\,kpc.  It is
unusual in that it had
a radio (Taylor, Frail \& Kulkarni 
1999) but no optical afterglow.
Possibly this is the example of a star-forming galaxy with a dense core
in which the GRB occurred, with the optical afterglow missing
due to internal extinction. The optical light from the host
galaxy would then trace the
star formation happening at low optical depth.  The 8.4-GHz radio flux
is less than 40\,$\mu$Jy 
(Taylor et al.~1999). If this is a
submillimetre galaxy, it is thus likely to be a low-luminosity one
unless it is at very high redshift.

\subsubsection{GRB\,971214}

{\it HST} images show a compact ($r_h=1.1$\,kpc), possibly merging 
$V_{\rm AB} \simeq 26.5$ 
host galaxy 
(Odewahn et al.~1998).  
The optical colours of the burst suggest substantial internal
extinction (Chary et al.~2002), and so it would be a good candidate
for being a submillimetre galaxy.
The radio imaging of this burst (Frail \& Kulkarni 1998) is moderately
deep; if this is a submillimetre galaxy, its bolometric luminosity is
unlikely to be more than twice that of the host galaxy of GRB 980703.

\subsubsection{GRB\,970828}

Like GRB\,990506, this burst had a radio, but no optical
afterglow. 
A burst occurring in such an object would
be a good candidate for having its optical (but not radio)
afterglow extinguished due to dust extinction.
This possibility is discussed in detail by Djorgovski et al.~(2001b), 
whose deep ground-based optical images show 
that the host appears to be a merging/interacting system.
They note that {\it HST} images 
should soon provide important information about the
morphology of the environment of
this optically-dark burst.

\subsubsection{GRB\,970508}

{\it HST}-STIS images of this host galaxy are presented
by Pian et al.~(1998), but the afterglow still dominates the optical
light.
A subsequent image (Fruchter et al.~1998) reveals the
host galaxy to be a compact, undisturbed galaxy
with $r_h=
0.8$\,kpc. 

\subsection{Compiling
a sample of low-luminosity submillimetre galaxies}

An obvious question is: how can one compile a sample of
low-luminosity submillimetre
galaxies.  This is an important exercise in that
it could reveal the bulk of cosmic star formation activity
taking place in the Universe.

An 850-$\mu$m detection of a GRB host galaxy
is the most direct method of establishing whether or
not it is a submillimetre galaxy.  The drawbacks of
this method are that integration times to get down to
the 2-mJy SCUBA confusion limit are long, and 
only the more luminous submillimetre galaxies, not
necessarily the ones that generate most cosmic star
formation, can be detected.
 
Radio observations of GRB host galaxies
offer another powerful method.  
From the calibration for the Carilli--Yun radio--submm spectral index 
of Dunne, Clements \& Eales~(2000), if we assume
a radio spectral slope $\alpha=0.32$ (Berger et al.~2001) and
a dust emissivity index $\beta=1.5$,
then the 1.4-GHz flux of a submillimetre galaxy at redshift $z<3$
\begin{equation}
S_{1.4} \approx 900 \, (1+z)^{-3.8}
\, \left( {{S_{850}}\over{2 \, {\rm mJy}}}\right)\, \mu{\rm Jy}.
\end{equation}
The normalization of equation (1)
is quite uncertain, especially at high redshifts.
For example, if we use
the conversion factor of Carilli \& Yun (2000), 
1.4-GHz fluxes a factor of 2
lower are predicted.
In our picture,
many GRB host galaxies will have 1.4-GHz radio fluxes of several,
maybe tens of,
microjanskies.  In other words, galaxies like the host of
GRB 980703 would be common and deep radio surveys of GRB host galaxies should
reveal several more.

If half of the star formation in the Universe happens in submillimetre
sources with $S_{850}$ between 0.4 and 4 mJy,
at redshifts of about 2, then we would expect
about half of GRB host galaxies to have 1.4-GHz fluxes $S_{1.4} >
3\,\mu$Jy.  If we restrict a survey to GRB hosts with known redshifts
$z<1$, then we would expect about
half of GRB host galaxies to have $S_{1.4} > 
13\,\mu$Jy.
For comparison, the host galaxy of GRB\,980703 has
$S_{1.4}=68\,\mu$Jy (Berger et al.~2001), the high flux reflecting  
the relatively low redshift $z=0.97$ of this burst.
At the VLA, \footnote{\tt http://www.aoc.nrao.edu/vla/html/VLAhome.shtml} a
1-$\sigma$ sensitivity $S_{1.4} \simeq 5\,\mu$Jy 
can be achieved in 24 hours.
From equation (1), $S_{1.4}$
is a steeply decreasing
function of redshift.  This suggests that low-redshift GRBs like
GRB 990712 ($z=0.43$) would be
the best targets for deep radio imaging.

Imaging of GRB host galaxies at a wavelength
70 $\mu$m with {\it SIRTF}
\footnote{\tt http://sirtf.caltech.edu/}
may help too,
particularly for the lower-redshift objects.
Direct submillimetre surveys with ALMA will uncover such objects,
and we will no longer
need to rely on using GRBs as beacons. ALMA will also be able to detect 
a 1-mJy host galaxy in less than a minute, and so can be used to image 
{\it all} known GRB positions in a matter of a few weeks of observations. 

Should submillimetre host galaxies not turn up in any of
the samples described in this section, this would also be
an interesting result.  One possibility could be that
GRBs avoid submillimetre galaxies because the environments 
there are not condusive to the physical processes
responsible for generating GRBs (metallicity, likely
to be high in submillimetre-luminous galaxies, could play
a role for example; see Ramirez-Ruiz, Lazzati \& Blain 2002).  
Alternatively it could mean that
our estimates of $f_{\rm submm}$ are too high,
possibly because the assumptions made in modeling the
luminosity evolution of submillimetre galaxies are  
flawed, for example if high-redshift submillimetre 
galaxies actually have dust temperatures systematically less than 
the 40\,K that seems to be typical. 

\subsection{Obscuration in low-luminosity submillimetre GRB host galaxies} 

Host galaxies like that of GRB\,980703 
appear to have star formation that is only
lightly obscured, intermediate between
Class-2 SCUBA galaxies 
(which would have brighter $R$ magnitudes) and ULIG cores (which would
extinguish optical afterglows).

The occasions where GRBs {\it did} appear to happen in dense cores were
the two occasions (GRB 970828 and 
GRB 990506) where radio but not optical
afterglows were detected.  That this has been
observed twice suggests that star
formation in dense cores at high redshifts is not as rare as
at low redshifts, where the contribution of ULIGs 
to the star formation
rate at $z=0$ (Gallego et al.~1995) is negligible. 

But the majority of high-redshift submillimetre galaxies, if the hosts
of GRB\,010222, GRB\,000418 and GRB\,980703 are at all 
representative, are probably
{\it not} of this type.  Their gas and dust  
are likely to be less concentrated.  Such a scenario is required
if most of the stars currently seen in local galaxies formed in
submillimetre galaxies (Trentham 2000).  These
submillimetre galaxies would be intermediate in terms of
structural parameters between Lyman-break galaxies and ULIGs,
possibly of the type hypothesized 
to exist on the boundary of detectability in both the optical 
and submillimetre wavebands by Adelberger \& Steidel (2000). 

\section*{Acknowledgements}
We thank 
Vicki Barnard, Fiona Harrison, 
Davide Lazzati, Priya Natarajan, Martin Rees, Dave Sanders,
Nial Tanvir and Bram Venemans for helpful conversations, and the 
anonymous referee for useful comments on the manuscript.  
We thank Jeff Goldader for making the data and results from
the $HST$-STIS imaging program of local infrared-luminous galaxies
available and Ian Smail for providing the SCUBA source catalogue paper
prior to publication.
ERR acknowledges support from CONACYT, SEP and the ORS foundation.    
AWB acknowledges the Raymond \& Beverly
Sackler Foundation for financial support at the IoA.
This work has used $HST$ data provided by 
the Survey of the Host Galaxies of Gamma-Ray Bursts
({\tt http://www.ifa.au.dk/$^{\sim}$hst/grb$_{-}$hosts/data/
index.html}).

\vskip 8pt

\end{document}